\documentclass[english,aps,prl,graphicx,twocolumn,showpacs]{revtex4}
\usepackage{ae}
\usepackage{aecompl}
\usepackage[T1]{fontenc}
\usepackage[latin1]{inputenc}
\usepackage{graphicx}

\makeatletter

\providecommand{\tabularnewline}{\\}

\makeatletter

\makeatletter

\usepackage{float}

\makeatletter

\usepackage{float}

\usepackage{epsfig}

\makeatother

\makeatother

\makeatother

\usepackage{babel}
\makeatother
\begin{document}

\title{Fractional charge perspective on the band-gap in density-functional
theory }

\author{Aron J. Cohen, Paula Mori-S\'{a}nchez, and Weitao Yang }

\affiliation{Department of Chemistry, Duke University, Durham, North Carolina
27708}

\date{\today{}}

\begin{abstract}
The calculation of the band-gap by density-functional theory (DFT) methods is examined
by considering the behavior of the energy as a function of number
of electrons. It is found that the incorrect band-gap prediction with most approximate
functionals originates mainly from errors in describing systems
with fractional charges. Formulas for the energy derivatives with
respect to number of electrons are derived which clarify the role
of optimized effective potentials in prediction of the band-gap.
Calculations with a recent functional that has much improved behavior
for fractional charges give a good prediction of the energy gap and
also $\varepsilon_{{\rm homo}}\simeq-I$ for finite systems. Our results
indicate it is possible, within DFT, to have a functional whose eigenvalues
or derivatives accurately predict the band-gap.

\end{abstract}
\maketitle

One of the many important uses of density-functional theory (DFT) is the
calculation of the band-structure which has many applications throughout physics,
for example in semiconductors, electron transport and reactions at surfaces.
The first step in achieving accuracy in the band structure is to understand the
band-gap which standard functionals have long been known to systematically underestimate
by as much as $\sim$50\%.
Recent efforts have focused on use of
the optimized effective potential (OEP) method, which can often give
an improvement in the prediction of band-gaps for small-gap semiconductors,
but has problems with wider gap semiconductors and insulators \cite{Stadele972089,Gruning06154108,Gruning06161103}.
In many cases it has proved necessary to move to
the quasi-particle GW theory to calculate the band-gap of solids accurately
\cite{Godby8810159}.
Conventionally, the band-gap problem, has been
related to the so-called {}``derivative discontinuity\char`\"{} in
the exchange correlation potential: even with an accurate Kohn-Sham
potential, the energy-gap is still different from the true gap by
an amount of the derivative discontinuity \cite{Perdew831884,Sham831888}.
This perspective, however, does not offer the understanding or the
mechanism needed for making progress for band-gap prediction with
DFT.

In this Letter, a new perspective is offered: the band-gap problem
is shown to be related to the behavior of approximate density functionals
for fractional numbers of electrons, an issue which has drawn considerable
recent interest \cite{Zhang982604,Morisanchez06201102,Ruzsinszky06104112}.
This enables us to understand the problem with band-gap calculations
and offers ideas to develop
functionals which predict the band-gap correctly. Examples will be
given for molecules where the energy gap can be compared with explicit
calculations of systems with fractional charges.

The fundamental band-gap for an $N$-electron system in an external
potential $v({\normalcolor \mathbf{r}})$, is given by \begin{eqnarray}
E_{g} & = & \big[E_{v}(N-1)-E_{v}(N)\big]-\big[E_{v}(N)-E_{v}(N+1)\big]\nonumber \\
 & = & I-A,\label{ImA}\end{eqnarray}
where $E_{v}(N)$ is the ground-state energy of the $N$ particle
system and $I$ is the ionization energy and $A$ is the electron
affinity. For a system with a fractional number of electrons $N+\delta N$,
with $0<\delta N<1$, it has been shown that the energy is a straight
line connecting the total energies at integer numbers of electrons;
namely, $E_{v}(N+\delta N)=\delta NE_{v}(N)+(1-\delta N)E_{v}(N+1)$
\cite{Perdew821691,Yang005172}. 
This linear relation means that the energy gap in Eq. (\ref{ImA})
can be given by the derivative difference \begin{eqnarray}
E_{g}^{\mathrm{der}} & = & \lim_{\delta N\rightarrow0}\left\{ \left.\frac{\partial E_{v}}{\partial N}\right|_{N+\delta N}-\left.\frac{\partial E_{v}}{\partial N}\right|_{N-\delta N}\right\} \label{sline}\end{eqnarray}
 If we substitute in the DFT total energy expression $E_{v}=T_{s}[\rho]+V_{{\rm ext}}[\rho]+J[\rho]+E_{xc}[\rho]$,
 we have \begin{equation}
E_{g}=\varepsilon_{{\rm lumo}}-\varepsilon_{{\rm homo}}+\Delta_{xc}=E_{g}^{{\rm KS}}+\Delta_{xc}\label{delta}\end{equation}
 where the $E_{g}^{{\rm KS}}$ is the gap in a Kohn-Sham calculation
and the $\Delta_{xc}$ is the derivative discontinuity \cite{Perdew831884,Sham831888}.

In this work we identify the problem with calculations using approximate
density functionals by considering the basic assumption in Eq. (\ref{sline}),
that the energy at $N+1$ and at $N-1$ can be given simply from the
derivatives at $N$. This is true for exact DFT but it may or may
not be true for approximate functionals. The key is to investigate
the behavior of the total energy 
as a function of numbers of electrons. To do this we consider a non-interacting
ground state reference system where we allow the occupation numbers
of the orbitals to vary the number of electrons smoothly; the first-order
reduced density matrix of the reference system is given by \begin{equation}
\rho_{s}(\mathbf{r},\mathbf{r}^{\prime})=\sum_{i}n_{i}\phi_{i}(\mathbf{r})\phi_{i}^{*}(\mathbf{r}^{\prime})\label{eq:frac_KS_density}\end{equation}
 where $n_{i}=1$ for $i<{\rm f}$, $n_{i}=\delta N$ for $i={\rm f}$$,$
and $n_{i}=0,$ for $i>{\rm f}$, and ${\rm f}$ is the index for
the frontier orbital. The electron density is just its diagonal $\rho_{s}({\bf \textbf{r}})=\rho_{s}({\bf \textbf{r},\textbf{r}})$.
The behavior of three qualitatively different exchange-correlation
functionals: the local density approximation (LDA), Hartree-Fock (HF)
and MCY3 \cite{Cohen07191109} is shown in Fig. 1. MCY3 was constructed
as a functional of $\rho_{s}(\mathbf{r},\mathbf{r}^{\prime})$ to
give a much improved description of fractional numbers of electrons
and we can see that it gives a straight line interpolation between
the integers. The interpolation from the other functionals is far
from linear, HF curves in a concave manner and LDA in a very convex
manner. There are two main problems with approximate density functionals:
First, they can have a remarkably different behavior from the exact
functional in fractional charge systems, compared to their behavior
for the corresponding integer charge systems. Second, the error in
the integer charge systems can also be significant. The combined effects
lead to the error in the band-gap prediction from derivative information.

\begin{figure}[!t]
 \includegraphics[width=0.5\textwidth]{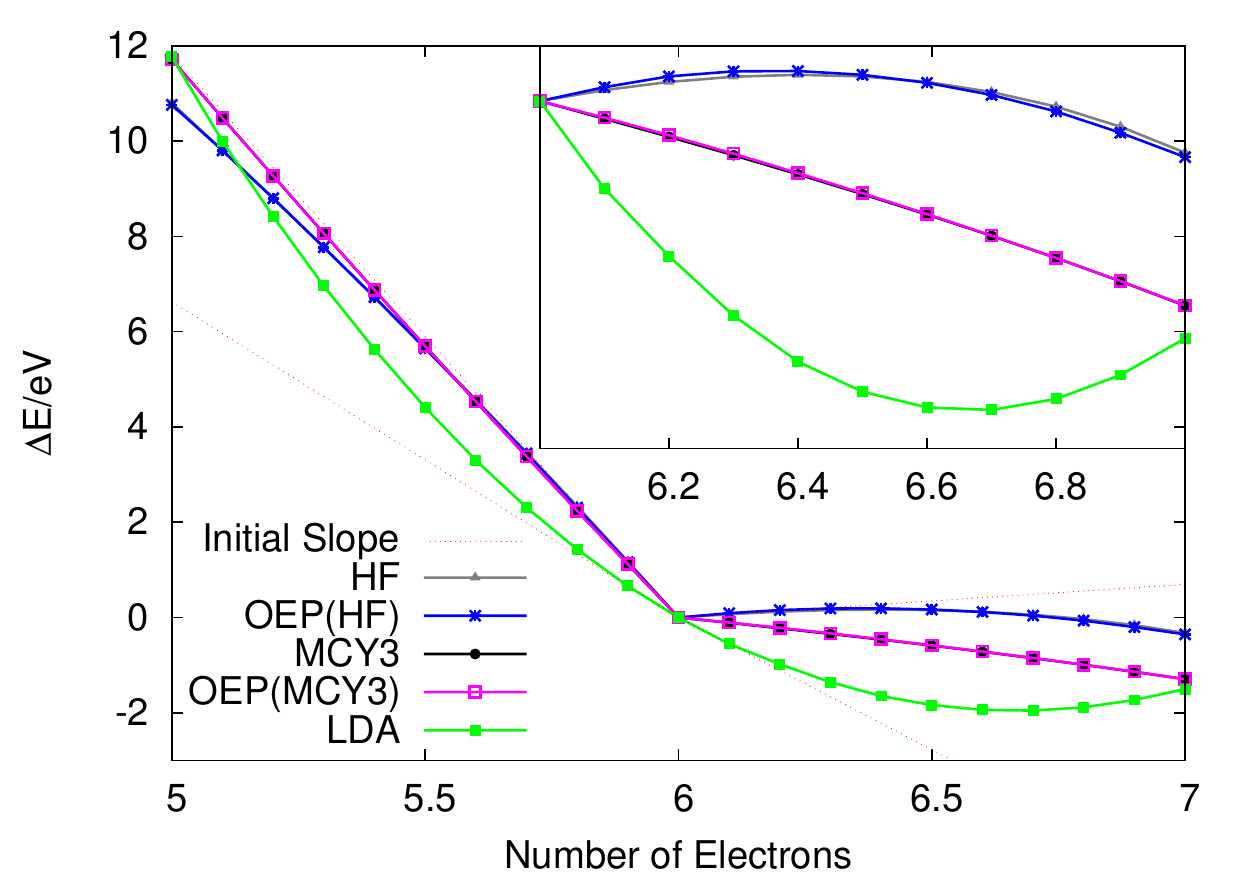} 
\caption{Behavior of the energy of the carbon atom with number of electron with fractional charges.
for several different functionals minimized with and without OEP.
Dotted line follows the initial slope for the non-straight functionals.
Inset shows $6<N<7$ range in more detail.}
\end{figure}

For molecules, LDA has a very reasonable description of the integer
values ($I$ and $A$ are given well) but a much worse description
in between the integers. The use of the first derivatives for LDA
will clearly not give the $I$ and $A$ from the integer calculations.
Because of the convexity of the curve, LDA will give too small a value
for $I$ and too large a value for $A$, meaning that the band-gap
$I-A$ will be too small as shown by the dotted lines in Fig 1. The
case for HF is very different as the integer values are not as good
because of the lack of correlation, and also it curves in a concave
manner. These two errors cancel each other to some extent in the prediction
of $I$ but add together in the calculation of $A$. For HF the value
of $I$ will be about right and $A$ too small, meaning that the band-gap
$I-A$ will be too large as shown in Fig 1. For MCY3, as it is very
straight, the use of the derivatives will give a prediction very similar
to the integer calculations for $I$ and $A$, and the band-gap $I-A$
should be accurate as the integer values are good.
\begin{figure}[!t]
 \includegraphics[width=0.5\textwidth]{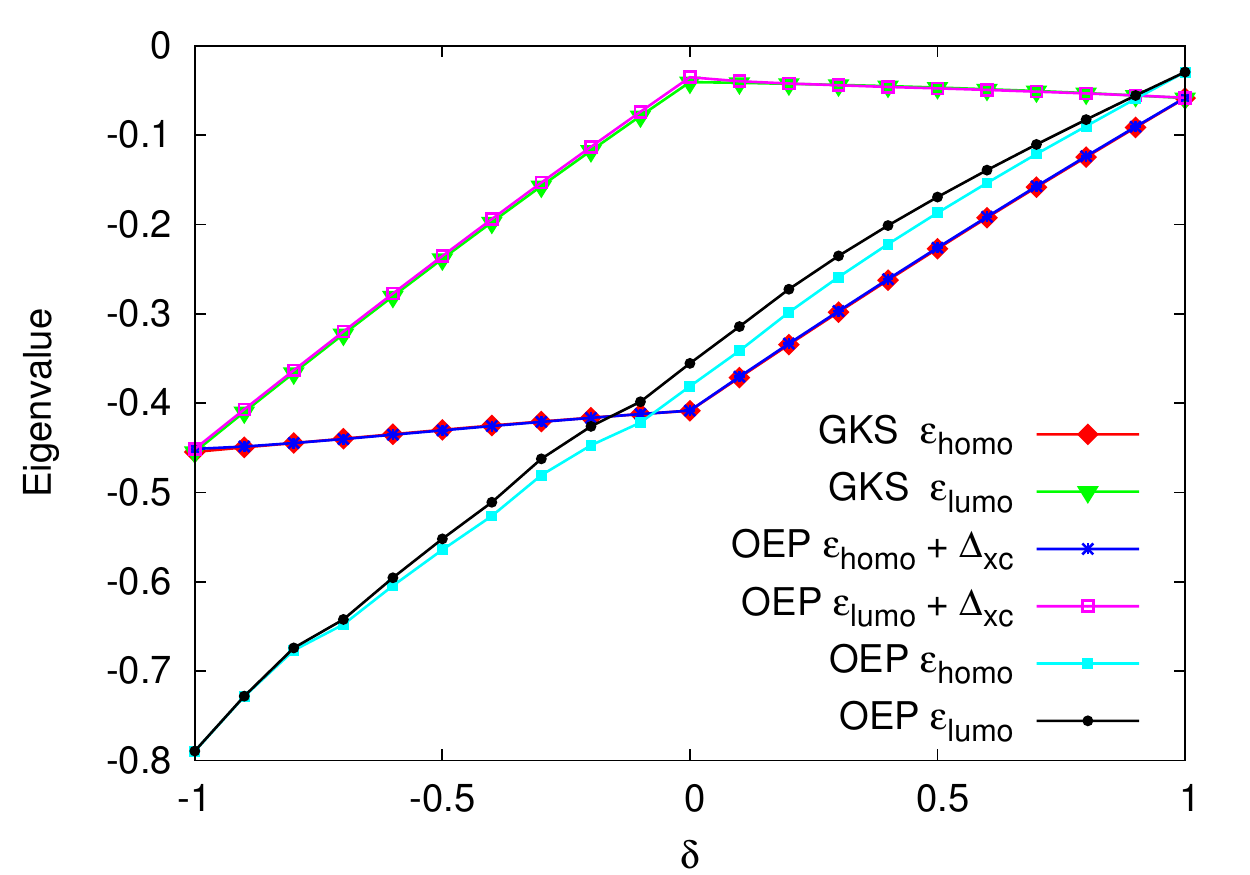} 
\caption{Comparison of MCY3 eigenvalues from GKS and OEP calculations and
also including $\Delta_{xc}$.}
\end{figure}
\begin{table*}[!t]

\caption{Comparison of $\varepsilon_{{\rm f}}$ against corresponding experimental
numbers for LDA, HF and MCY3 for more details see \cite{Cohen07PRLSup}\label{table1}}

\begin{tabular}{lcccc|cccc|cccc}
\hline 
\hline 
&
\multicolumn{3}{c}{$\varepsilon_{{\rm lumo}}-\varepsilon_{{\rm homo}}$}&
$I-A$ &
\multicolumn{3}{c}{$-\varepsilon_{{\rm homo}}$}&
$I$ &
\multicolumn{3}{c}{$-\varepsilon_{{\rm lumo}}$}&
$A$\tabularnewline
Mol &
LDA &
HF &
\multicolumn{1}{c}{MCY3}&
Expt &
LDA &
HF&
\multicolumn{1}{c}{MCY3}&
Expt &
LDA &
HF &
\multicolumn{1}{c}{MCY3}&
Expt\tabularnewline
\hline 
C &
0.08 &
12.76 &
10.03 &
10.00 &
6.09 &
11.94 &
11.12 &
11.27 &
6.01 &
-0.82 &
1.09 &
1.27 \tabularnewline
O &
0.23 &
16.80 &
11.57 &
12.16 &
7.28 &
14.11 &
13.01 &
13.62 &
7.05 &
-2.70 &
1.44 &
1.46 \tabularnewline
F$_{2}$ &
3.38 &
20.44 &
13.75 &
14.40 &
9.53 &
18.13 &
15.17 &
15.70 &
6.15 &
-2.31 &
1.42 &
1.30 \tabularnewline
OH &
0.10 &
16.56 &
11.23 &
11.40 &
7.21 &
13.90 &
12.70 &
13.20 &
7.11 &
-2.67 &
1.47 &
1.80 \tabularnewline
MAE &
4.678 &
2.082 &
0.319 &
&
5.059 &
0.990 &
0.484&
&
4.295 &
3.328 &
0.488&
\tabularnewline
\hline 
\multicolumn{13}{l}{Errors from the explicit calculation of $I$ and $A$ \cite{Cohen07PRLSup}} 
\tabularnewline
MAE &
0.920 &
1.248 &
0.789 &
&
0.659 &
1.235 &
0.505&
&
0.328 &
1.655 &
0.475&
\tabularnewline
\hline
\hline 
\end{tabular}
\end{table*}
For functionals that have a linear behavior for fractional charge
systems (e.g. MCY3 and the exact functional) the initial derivative
is all that is needed to calculate $I$ and $A$ and the
band-gap as in Eq. (\ref{sline}). Hence, we consider analytic expressions
for $\partial E_{v}/\partial N$ 
for some families of exchange-correlation functionals. The main ideas
and results are summarized here, with further details presented in
the supplementary material \cite{Cohen07PRLSup}.

In the fractional charge non-interacting system, Eq. (\ref{eq:frac_KS_density}),
the orbitals are the eigenstates of an one-electron local potential
$v_{s}({\bf \textbf{r}})$ \begin{equation}
\left(-\frac{1}{2}\nabla^{2}+v_{s}\right)\left|\phi_{i}\right\rangle =\varepsilon_{i}\left|\phi_{i}\right\rangle ,\label{eq:OEPEigen}\end{equation}
 or a non-local potential $v_{s}^{NL}({\bf \textbf{r,}}{\bf \textbf{r}}^{\prime})$.
\begin{equation}
\left(-\frac{1}{2}\nabla^{2}+v_{s}^{NL}\right)\left|\phi_{i}\right\rangle =\varepsilon_{i}^{GKS}\left|\phi_{i}\right\rangle .\label{eq:OEPEigen_NL}\end{equation}
 The former is the original Kohn-Sham (KS) reference system and the
latter has been called the Hartree-Fock-Kohn-Sham (HFKS) \cite{Parr89}
or the generalized Kohn-Sham (GKS) method \cite{Seidl963764}.

For the Kohn-Sham reference system with local potential $v_{s}({\bf \textbf{r}})$,
we here use the potential-functional formulation \cite{Yang04146404}.
The electron density $\rho_{s}({\bf \textbf{r}})$ can be represented
as the set of orbitals and occupation numbers $\left\{ \phi_{i},n_{i}\right\} $,
or equivalently as the local KS potential and total particle number
$\left\{ v_{s}({\bf \textbf{r}}),N\right\} $. Thus the total energy
functional, formally in term of density as $E_{v}[\rho_{s}({\bf \textbf{r}})]$,
can be expressed as $E_{v}[v_{s}({\bf \textbf{r}}),N]$. The ground
state energy is the minimum of the KS energy functional, expressed
(explicitly or implicitly) in terms of the local potential $v_{s}(\mathbf{{\bf \textbf{r}}})$: $E_{v}(N)=\min_{v_{s}}E_{v}[v_{s},N]=E_{v}[v_{s}^{gs},N],$
where the minimizer $v_{s}^{gs}$ is the optimized effective potential
(OEP), as established recently \cite{Yang04146404}. The variational
nature of $v_{s}^{gs}$ simplifies the calculation of the derivative:$\frac{\partial E_{v}(N)}{\partial N}=\left.\frac{\partial E_{v}[v_{s}^{gs},N]}{\partial N}\right|_{v_{s}^{gs}}$.
Consider a change in the total number of electrons $N=N_{0}+\delta N,\ $
where $N_{0}$ is an integer and $|\delta N|<1$. At fixed $v_{s}^{{\rm gs}}$,
all the orbitals $\left\{ \phi_{i}^{v_{s}^{gs}}\right\} $, as its
eigenstates, are fixed. Since $\rho_{s}({\bf r})$ is the ground state
density of the reference potential $v_{s}^{{\rm gs}}$, only the frontier
level occupation $n_{{\rm f}}$ is allowed to change $\delta N=\delta n_{{\rm f}}$,
thus \begin{equation}
\frac{\partial E_{v}(N)}{\partial N}=\left(\frac{\partial E_{v}[\{ \phi_{i}^{v_{s}^{gs}},n_{i}\} ]}{\partial n_{{\rm f}}}\right)_{\left\{ \phi_{i}^{v_{s}^{gs}}\right\} },\label{eq:dEdN_2}\end{equation}
 where the frontier orbital is either the LUMO, $n_{{\rm f}}=n_{{\rm lumo}},{\rm if}\;\delta N>0$
, or the HOMO, $n_{{\rm f}}=n_{{\rm homo}},{\rm if}\;\delta N<0.$
We consider three
cases for which the analytic derivatives can be obtained \cite{Cohen07PRLSup}:

\textbf{Case A}: $E_{xc}[\rho_{s}(\mathbf{r})]$, an explicit functional
of $\rho_{s}$ (e.g. LDA or GGA): 
\begin{equation}
\frac{\partial E_{v}(N)}{\partial N}=\varepsilon_{f}\end{equation}
 where $\varepsilon_{{\rm f}}$ is the KS eigenvalue for the frontier
orbital in the local potential $v_{s}(\mathbf{r})=v({\bf \textbf{r}})+v_{J}(\mathbf{\mathbf{\mathbf{r}}})+v_{xc}(\mathbf{\mathbf{\mathbf{r}}})$.
This is exactly the combination of Eq. (\ref{eq:dEdN_2}) with Janak's
theorem for $n_{{\rm f}}$ \cite{Janak787165}.

\textbf{Case B}: $E_{xc}[\rho_{s}({\bf \textbf{r}},{\bf \textbf{r}}^{\prime})]$,
a functional of the first order density matrix minimized with a local
potential as in Eq. (\ref{eq:OEPEigen}) (e.g. OEP exact exchange,
EXX). 
\begin{equation}
\frac{\partial E_{v}(N)}{\partial N}=\varepsilon_{{\rm f}}+\langle\phi_{{\rm f}}|v+v_{J}+v_{xc}^{NL}-v_{s}|\phi_{{\rm f}}\rangle,\label{eq:derivativeOEP_2}\end{equation}
where the non-local potential $v_{xc}^{NL}({\bf \mathbf{r},\mathbf{r}'})=\frac{\delta E_{xc}[\rho_{s}({\bf \mathbf{r},\mathbf{r}'})]}{\delta\rho_{s}({\bf \mathbf{r},\mathbf{r}'})}$.
Eq. (\ref{eq:derivativeOEP_2}) is a key result, showing that for
general orbital functionals, $\frac{\partial E_{v}}{\partial N}$
is not given by the frontier OEP eigenvalue, $\varepsilon_{{\rm f}}$
, but with a correction term $\Delta_{xc}^{\rm f}$. 
This general result agrees with \cite{Stadele972089} in the case
of exact exchange, and is related to the results of \cite{Casida994694}
from the self energy.

\textbf{Case C}: $E_{xc}[\rho_{s}({\bf \textbf{r}},{\bf \textbf{r}}^{\prime})]$,
with the energy minimized with respect to the orbitals $\phi_{i}$
(e.g. HF ):
\begin{equation}
\frac{\partial E_{v}(N)}{\partial N}=\varepsilon_{{\rm f}}^{GKS}\label{OEP}\end{equation}
 where $\varepsilon_{{\rm f}}^{GKS}$is the eigenvalue of the frontier
orbital of the non-local potential $v+v_{J}+v_{xc}^{NL}({\bf \mathbf{r},\mathbf{r}'})$,
as in Eq. (\ref{eq:OEPEigen_NL}).

All three cases can be unified in the expression 
\begin{equation}
\frac{\partial E_{v}(N)}{\partial N}=\langle\phi_{{\rm f}}|H_{{\rm eff}}|\phi_{{\rm f}}\rangle,\label{eq:dEdN_abc}\end{equation}
 where $H_{{\rm eff}}=-\frac{1}{2}\nabla^{2}+v+v_{J}+v_{xc}(\mathbf{r})$
for Case A where $E_{xc}=E_{xc}[\rho(\mathbf{r})]$, and $H_{{\rm eff}}=-\frac{1}{2}\nabla^{2}+v+v_{J}+v_{xc}^{NL}({\bf \mathbf{r},\mathbf{r}'})$
for case B and C where $E_{xc}=E_{xc}[\rho_{s}({\bf \textbf{r}}^{\prime},{\bf \textbf{r}})]$,
which are often called orbital functionals. In cases A and C, $\frac{\partial E_{v}(N)}{\partial N}$
is equal to the corresponding eigenvalue, but not in case B. The only
difference in between cases B and C is the
orbitals used to evaluate the overall expression.

Combining Eqs. (\ref{sline}) and (\ref{eq:dEdN_abc}) thus leads
to the general expression for the band-gap from derivatives for an
N-electron system: \begin{eqnarray}
E_{g}^{\mathrm{der}} & = & \langle\phi_{N+1}|H_{{\rm eff}}|\phi_{N+1}\rangle-\langle\phi_{N}|H_{{\rm eff}}|\phi_{N}\rangle.\label{sline-2}\end{eqnarray}

We consider a few illustrative atomic and molecular systems for which
we have performed self-consistent calculations using a cc-pVQZ basis
set in an modified version of CADPAC. We compare $-\varepsilon_{{\rm homo}}$
with the experimental $I$, $-\varepsilon_{{\rm lumo}}$ with the
experimental $A$ and also their corresponding differences. The results
for LDA, HF and MCY3 are given in Table 1. MCY3 gives very good
agreement between $-\varepsilon_{{\rm homo}}$ and $I$ which is to
be expected from its straight line behavior. We should emphasize that
this has not been seen before for calculations
with approximate exchange-correlation functionals. The error is relatively
small, 0.5eV, and is roughly similar in $I$, $A$ and also the difference
$I-A$. LDA eigenvalues have a large error, with a consistent underestimation
of $I$ by about 5 eV and overestimation of $A$ by about 4 eV and a 
poor predicition of the gap. 
LDA does well for the explicit calculation of the $N+1$ and $N-1$ systems,
it is just the use of the derivatives at $N$ that lead to large
errors, this is now clearly understood from the fractional charge
picture. The HF $-\varepsilon_{{\rm homo}}$ is often close to $I$, 
however there are larger errors for $A$ and also for
the gap. 
\begin{table}[!t]

\caption{Comparison of GKS and OEP eigenvalues, and $\partial E/\partial n_{\rm f}$
for HF and MCY3, see \cite{Cohen07PRLSup}.}

\begin{tabular}{lccccccccc}
\hline
\hline 
Mol &
&
MCY3 &
MCY3 &
MCY3 &
HF &
HF &
HF &
Expt &
\tabularnewline
&
&
\multicolumn{2}{c}{OEP}&
GKS &
\multicolumn{2}{c}{OEP}&
GKS &
&
\tabularnewline
&
&
$\varepsilon$ &
$\partial E/\partial n_{\rm f}$ &
$\varepsilon$ &
$\varepsilon$ &
$\partial E/\partial n_{\rm f}$ &
$\varepsilon$ &
&
\tabularnewline
\hline 
C &
I-A &
0.70 &
10.16 &
10.03 &
1.47 &
13.49 &
12.76 &
10.00 &
\tabularnewline
&
I &
10.58 &
11.11 &
11.12 &
11.97 &
11.94 &
11.94 &
11.27 &
\tabularnewline
&
A &
9.88 &
0.96 &
1.09 &
10.50 &
-1.55 &
-0.82 &
1.27 &
\tabularnewline
F$_{2}$ &
I-A &
4.06 &
13.74 &
13.75 &
5.62 &
20.49 &
20.44 &
14.40 &
\tabularnewline
&
I &
14.67 &
15.16 &
15.17 &
15.94 &
18.11 &
18.13 &
15.70 &
\tabularnewline
&
A &
10.61 &
1.42 &
1.42 &
10.32 &
-2.37 &
-2.31 &
1.30 &
\tabularnewline
\hline
\hline
\end{tabular}
\end{table}

The band-gap issue is  well understood for calculations with local
density functionals (case A) or orbitals functionals (case C). 
We now consider case B, orbital functionals in an OEP calculation,
using the Yang-Wu direct minimization method \cite{Yang02143002,Wu03627}.
In Fig. 1 the OEP minimized energy is remarkably similar to
the GKS minimized energy in both integer and fractional charge systems.
We would therefore expect $\partial E/\partial N_{N\pm\delta N}$ 
to be the same as the GKS derivatives.

Table 2 shows the eigenvalues from an OEP calculation using a Fermi-Amaldi
base potential, which has the correct asymptotic behavior. 
The asymptotic form of the potential has a large effect on the OEP eigenvalues,
but not on the eigenvalue differences or the energy derivatives.
The OEP(MCY3) $-\varepsilon_{{\rm homo}}\simeq I$
(as is proven for the exact functional \cite{Perdew9716021})
however the $E_{g}^{{\rm KS}}$ is much smaller than the exact gap.
The inclusion of $\Delta_{xc}^{\rm f}$ gives a much better
agreement between the derivatives and the GKS eigenvalues. This brings
us on to the nature of this term; it is only the difference
between a KS and GKS calculation and is needed to correctly give the
derivative at $N$. 
It does not, however, address the question of whether
the functional used for the calculation has the correct straight line
behavior for fractional numbers of electrons, which is the key question
in the evaluation of the band-gap.

Fig. 2 shows the behavior of the eigenvalues for carbon with
different numbers of electrons using MCY3. The GKS $\varepsilon_{{\rm homo}}$
for a fractional system is almost constant between integers due to 
the straight line behavior of MCY3.
The OEP $\epsilon_{\rm f}$ are markedly different to the
 GKS $\epsilon_{\rm f}$, but they become almost identical upon 
inclusion of $\Delta_{xc}^{\rm f}$,
which can be understood from comparing Eqs. (\ref{eq:derivativeOEP_2})
and (\ref{OEP}). 
The LUMO at $N-\delta N$ is connected to the HOMO at $N+\delta N$.
This is clear from Eq. (\ref{eq:dEdN_2}) as the change in the 
number of electrons, $N$, is only through the frontier occupation
numbers, $n_{{\rm f}}$, and the potential
and, therefore, eigenfunctions remain fixed, i.e. there is no mysterious
discontinuity in the eigenvalues. 

In conclusion we have carried out analysis and calculations on systems
with fractional numbers of electrons to gain understanding of the
band-gap problem in DFT. We show that the band-gap is only given by
the eigenvalue difference if the functional has the correct linear
behavior for systems with a fractional charge. We have recently developed
a functional with this linear behavior giving 
$-\varepsilon_{{\rm homo}}\simeq I$
and $-\varepsilon_{{\rm lumo}}\simeq A$ and a good prediction of the
band-gap in molecules. We have also considered OEP calculations, in
which the derivative of the energy with respect to number of electrons
is not given by the OEP eigenvalue.
When the derivative is correctly evaluated, it gives practically the
same as in GKS calculations. 
Our work thus provides the new insight: it is possible to have a functional
which gives the correct band-gap from the eigenvalues or derivative
information, so long as it has the correct fractional charge behavior
and accurate energies for integer systems. Such possible functionals 
include explicit functionals of the electron density $E_{xc}[\rho(\mathbf{r})]$.
We have only considered the explicit calculation of molecules but
the same ideas are undoubtedly of key importance in solids. 
The understanding gained in this Letter offers a new perspective
and way forward for accurate calculations of the band-gap in DFT.

Support from NSF is greatly appreciated.

\bibliographystyle{apsrev}

\begin{thebibliography}{21}
\expandafter\ifx\csname natexlab\endcsname\relax\def\natexlab#1{#1}\fi
\expandafter\ifx\csname bibnamefont\endcsname\relax
  \def\bibnamefont#1{#1}\fi
\expandafter\ifx\csname bibfnamefont\endcsname\relax
  \def\bibfnamefont#1{#1}\fi
\expandafter\ifx\csname citenamefont\endcsname\relax
  \def\citenamefont#1{#1}\fi
\expandafter\ifx\csname url\endcsname\relax
  \def\url#1{\texttt{#1}}\fi
\expandafter\ifx\csname urlprefix\endcsname\relax\def\urlprefix{URL }\fi
\providecommand{\bibinfo}[2]{#2}
\providecommand{\eprint}[2][]{\url{#2}}

\bibitem[{\citenamefont{St{\"a}dele et~al.}(1997)\citenamefont{St{\"a}dele,
  Majewski, Vogl, and G{\"o}rling}}]{Stadele972089}
\bibinfo{author}{\bibfnamefont{M.}~\bibnamefont{St{\"a}dele}},
  \bibinfo{author}{\bibfnamefont{J.~A.} \bibnamefont{Majewski}},
  \bibinfo{author}{\bibfnamefont{P.}~\bibnamefont{Vogl}}, \bibnamefont{and}
  \bibinfo{author}{\bibfnamefont{A.}~\bibnamefont{G{\"o}rling}},
  \bibinfo{journal}{Phys. Rev. Lett.} \textbf{\bibinfo{volume}{79}},
  \bibinfo{pages}{2089} (\bibinfo{year}{1997}).

\bibitem[{\citenamefont{Gruning
  et~al.}(2006{\natexlab{a}})\citenamefont{Gruning, Marini, and
  Rubio}}]{Gruning06154108}
\bibinfo{author}{\bibfnamefont{M.}~\bibnamefont{Gruning}},
  \bibinfo{author}{\bibfnamefont{A.}~\bibnamefont{Marini}}, \bibnamefont{and}
  \bibinfo{author}{\bibfnamefont{A.}~\bibnamefont{Rubio}}, \bibinfo{journal}{J.
  Chem. Phys.} \textbf{\bibinfo{volume}{124}}, \bibinfo{pages}{154108}
  (\bibinfo{year}{2006}{\natexlab{a}}).

\bibitem[{\citenamefont{Gruning
  et~al.}(2006{\natexlab{b}})\citenamefont{Gruning, Marini, and
  Rubio}}]{Gruning06161103}
\bibinfo{author}{\bibfnamefont{M.}~\bibnamefont{Gruning}},
  \bibinfo{author}{\bibfnamefont{A.}~\bibnamefont{Marini}}, \bibnamefont{and}
  \bibinfo{author}{\bibfnamefont{A.}~\bibnamefont{Rubio}},
  \bibinfo{journal}{Phys. Rev. B} \textbf{\bibinfo{volume}{74}},
  \bibinfo{pages}{161103(R)} (\bibinfo{year}{2006}{\natexlab{b}}).

\bibitem[{\citenamefont{Godby et~al.}(1988)\citenamefont{Godby, Schl\"uter, and
  Sham}}]{Godby8810159}
\bibinfo{author}{\bibfnamefont{R.~W.} \bibnamefont{Godby}},
  \bibinfo{author}{\bibfnamefont{M.}~\bibnamefont{Schl\"uter}},
  \bibnamefont{and} \bibinfo{author}{\bibfnamefont{L.~J.} \bibnamefont{Sham}},
  \bibinfo{journal}{Phys. Rev. B} \textbf{\bibinfo{volume}{37}},
  \bibinfo{pages}{10159} (\bibinfo{year}{1988}).

\bibitem[{\citenamefont{Perdew and Levy}(1983)}]{Perdew831884}
\bibinfo{author}{\bibfnamefont{J.~P.} \bibnamefont{Perdew}} \bibnamefont{and}
  \bibinfo{author}{\bibfnamefont{M.}~\bibnamefont{Levy}},
  \bibinfo{journal}{Phys. Rev. Lett.} \textbf{\bibinfo{volume}{51}},
  \bibinfo{pages}{1884} (\bibinfo{year}{1983}).

\bibitem[{\citenamefont{Sham and Schl\"uter}(1983)}]{Sham831888}
\bibinfo{author}{\bibfnamefont{L.~J.} \bibnamefont{Sham}} \bibnamefont{and}
  \bibinfo{author}{\bibfnamefont{M.}~\bibnamefont{Schl\"uter}},
  \bibinfo{journal}{Phys. Rev. Lett.} \textbf{\bibinfo{volume}{51}},
  \bibinfo{pages}{1888} (\bibinfo{year}{1983}).

\bibitem[{\citenamefont{Zhang and Yang}(1998)}]{Zhang982604}
\bibinfo{author}{\bibfnamefont{Y.}~\bibnamefont{Zhang}} \bibnamefont{and}
  \bibinfo{author}{\bibfnamefont{W.}~\bibnamefont{Yang}}, \bibinfo{journal}{J.
  Chem. Phys.} \textbf{\bibinfo{volume}{109}}, \bibinfo{pages}{2604}
  (\bibinfo{year}{1998}).

\bibitem[{\citenamefont{Mori-S\'{a}nchez
  et~al.}(2006)\citenamefont{Mori-S\'{a}nchez, Cohen, and
  Yang}}]{Morisanchez06201102}
\bibinfo{author}{\bibfnamefont{P.}~\bibnamefont{Mori-S\'{a}nchez}},
  \bibinfo{author}{\bibfnamefont{A.~J.} \bibnamefont{Cohen}}, \bibnamefont{and}
  \bibinfo{author}{\bibfnamefont{W.}~\bibnamefont{Yang}}, \bibinfo{journal}{J.
  Chem. Phys.} \textbf{\bibinfo{volume}{125}}, \bibinfo{pages}{201102}
  (\bibinfo{year}{2006}).

\bibitem[{\citenamefont{Ruzsinszky et~al.}(2006)\citenamefont{Ruzsinszky,
  Perdew, Csonka, I, Vydrov, and Scuseria}}]{Ruzsinszky06104112}
\bibinfo{author}{\bibfnamefont{A.}~\bibnamefont{Ruzsinszky}},
  \bibinfo{author}{\bibfnamefont{J.~P.} \bibnamefont{Perdew}},
  \bibinfo{author}{\bibfnamefont{G.~I.} \bibnamefont{Csonka}},
  \bibinfo{author}{\bibnamefont{I}}, \bibinfo{author}{\bibfnamefont{O.~A.}
  \bibnamefont{Vydrov}}, \bibnamefont{and}
  \bibinfo{author}{\bibfnamefont{G.~E.} \bibnamefont{Scuseria}},
  \bibinfo{journal}{J. Chem. Phys.} \textbf{\bibinfo{volume}{126}},
  \bibinfo{pages}{104102} (\bibinfo{year}{2006}).

\bibitem[{\citenamefont{Perdew et~al.}(1982)\citenamefont{Perdew, Parr, Levy,
  and {Balduz Jr.}}}]{Perdew821691}
\bibinfo{author}{\bibfnamefont{J.~P.} \bibnamefont{Perdew}},
  \bibinfo{author}{\bibfnamefont{R.~G.} \bibnamefont{Parr}},
  \bibinfo{author}{\bibfnamefont{M.}~\bibnamefont{Levy}}, \bibnamefont{and}
  \bibinfo{author}{\bibfnamefont{J.~L.} \bibnamefont{{Balduz Jr.}}},
  \bibinfo{journal}{Phys. Rev. Lett.} \textbf{\bibinfo{volume}{49}},
  \bibinfo{pages}{1691} (\bibinfo{year}{1982}).

\bibitem[{\citenamefont{Yang et~al.}(2000)\citenamefont{Yang, Zhang, and
  Ayers}}]{Yang005172}
\bibinfo{author}{\bibfnamefont{W.}~\bibnamefont{Yang}},
  \bibinfo{author}{\bibfnamefont{Y.}~\bibnamefont{Zhang}}, \bibnamefont{and}
  \bibinfo{author}{\bibfnamefont{P.}~\bibnamefont{Ayers}},
  \bibinfo{journal}{Phys. Rev. Lett.} \textbf{\bibinfo{volume}{84}},
  \bibinfo{pages}{5172} (\bibinfo{year}{2000}).

\bibitem[{\citenamefont{Cohen et~al.}(2007)\citenamefont{Cohen,
  Mori-S\'{a}nchez, and Yang}}]{Cohen07191109}
\bibinfo{author}{\bibfnamefont{A.~J.} \bibnamefont{Cohen}},
  \bibinfo{author}{\bibfnamefont{P.}~\bibnamefont{Mori-S\'{a}nchez}},
  \bibnamefont{and} \bibinfo{author}{\bibfnamefont{W.}~\bibnamefont{Yang}},
  \bibinfo{journal}{J. Chem. Phys.} \textbf{\bibinfo{volume}{126}},
  \bibinfo{pages}{191109} (\bibinfo{year}{2007}).

\bibitem[{Coh()}]{Cohen07PRLSup}
\bibinfo{note}{See EPAPS Document No. ? for details of the derivation of
  Equations (10)-(13) and additional calculations on molecules. This document
  can be reached via a direct link in the online article's HTML reference
  section or via the EPAPS homepage (http://www.aip.org/pubservs/epaps.html).}

\bibitem[{\citenamefont{Parr and Yang}(1989)}]{Parr89}
\bibinfo{author}{\bibfnamefont{R.~G.} \bibnamefont{Parr}} \bibnamefont{and}
  \bibinfo{author}{\bibfnamefont{W.}~\bibnamefont{Yang}},
  \emph{\bibinfo{title}{Density-Functional Theory of Atoms and Molecules}}
  (\bibinfo{publisher}{Oxford University Press}, \bibinfo{address}{New York},
  \bibinfo{year}{1989}).

\bibitem[{\citenamefont{Seidl et~al.}(1996)\citenamefont{Seidl, G\"orling,
  Vogl, Majewski, and Levy}}]{Seidl963764}
\bibinfo{author}{\bibfnamefont{A.}~\bibnamefont{Seidl}},
  \bibinfo{author}{\bibfnamefont{A.}~\bibnamefont{G\"orling}},
  \bibinfo{author}{\bibfnamefont{P.}~\bibnamefont{Vogl}},
  \bibinfo{author}{\bibfnamefont{J.~A.} \bibnamefont{Majewski}},
  \bibnamefont{and} \bibinfo{author}{\bibfnamefont{M.}~\bibnamefont{Levy}},
  \bibinfo{journal}{Phys. Rev. B} \textbf{\bibinfo{volume}{53}},
  \bibinfo{pages}{3764} (\bibinfo{year}{1996}).

\bibitem[{\citenamefont{Yang et~al.}(2004)\citenamefont{Yang, Ayers, and
  Wu}}]{Yang04146404}
\bibinfo{author}{\bibfnamefont{W.}~\bibnamefont{Yang}},
  \bibinfo{author}{\bibfnamefont{P.~W.} \bibnamefont{Ayers}}, \bibnamefont{and}
  \bibinfo{author}{\bibfnamefont{Q.}~\bibnamefont{Wu}}, \bibinfo{journal}{Phys.
  Rev. Lett.} \textbf{\bibinfo{volume}{92}}, \bibinfo{pages}{146404}
  (\bibinfo{year}{2004}).

\bibitem[{\citenamefont{Janak}(1978)}]{Janak787165}
\bibinfo{author}{\bibfnamefont{J.~F.} \bibnamefont{Janak}},
  \bibinfo{journal}{Phys. Rev. B} \textbf{\bibinfo{volume}{18}},
  \bibinfo{pages}{7165} (\bibinfo{year}{1978}).

\bibitem[{\citenamefont{Casida}(1999)}]{Casida994694}
\bibinfo{author}{\bibfnamefont{M.~E.} \bibnamefont{Casida}},
  \bibinfo{journal}{Phys. Rev. B} \textbf{\bibinfo{volume}{59}},
  \bibinfo{pages}{4694} (\bibinfo{year}{1999}).

\bibitem[{\citenamefont{Yang and Wu}(2002)}]{Yang02143002}
\bibinfo{author}{\bibfnamefont{W.}~\bibnamefont{Yang}} \bibnamefont{and}
  \bibinfo{author}{\bibfnamefont{Q.}~\bibnamefont{Wu}}, \bibinfo{journal}{Phys.
  Rev. Lett.} \textbf{\bibinfo{volume}{89}}, \bibinfo{pages}{143002/1}
  (\bibinfo{year}{2002}).

\bibitem[{\citenamefont{Wu and Yang}(2003)}]{Wu03627}
\bibinfo{author}{\bibfnamefont{Q.}~\bibnamefont{Wu}} \bibnamefont{and}
  \bibinfo{author}{\bibfnamefont{W.}~\bibnamefont{Yang}}, \bibinfo{journal}{J.
  Theo. Comp. Chem.} \textbf{\bibinfo{volume}{2}}, \bibinfo{pages}{627}
  (\bibinfo{year}{2003}).

\bibitem[{\citenamefont{Perdew and Levy}(1997)}]{Perdew9716021}
\bibinfo{author}{\bibfnamefont{J.~P.} \bibnamefont{Perdew}} \bibnamefont{and}
  \bibinfo{author}{\bibfnamefont{M.}~\bibnamefont{Levy}},
  \bibinfo{journal}{Phys. Rev. B} \textbf{\bibinfo{volume}{56}},
  \bibinfo{pages}{16021} (\bibinfo{year}{1997}).

\end{thebibliography}

\end{document}